\documentclass[
  reprint,           
  aps,
  prl,
  amsmath,
  amssymb,
  superscriptaddress,
  nofootinbib,
  floatfix
]{revtex4-2}

\usepackage{mathtools} 
\usepackage{bm}
\usepackage{dsfont}
\usepackage{graphicx}
\usepackage{pifont}
\usepackage{textcomp}
\usepackage{gensymb}
\usepackage{accents}
\usepackage{upgreek}
\usepackage{array}
\usepackage{hhline}
\usepackage{tabularx}
\usepackage{color}
\usepackage{multirow}
\usepackage[hidelinks]{hyperref}

\usepackage{ulem}
\usepackage[version=3]{mhchem} 
\usepackage{fixmath}

\makeatletter
\newcommand*{\supplementarystart}{%
  \close@column@grid%
  \clearpage%
  \onecolumngrid%
  \setcounter{enumiv}{0} 
  \setcounter{equation}{0} 
  \setcounter{figure}{0} 
  \setcounter{table}{0} 
  \setcounter{page}{1}
  \c@secnumdepth=4
  \renewcommand{\theequation}{s\arabic{equation}} 
  \renewcommand{\bibnumfmt}[1]{[s##1]} 
  \renewcommand{\cite}[1]{{[}\onlinecite{##1}{]}}
  \renewcommand{\thefigure}{s\arabic{figure}}
  \renewcommand{\thetable}{s\Roman{table}}
  \renewcommand{\thepage}{s\arabic{page}}
}
\makeatother

\newcommand{\pa}{\partial}

\newcommand{\be}{\begin{equation}}
\newcommand{\e}{\end{equation}}
\newcommand{\beml}{\begin{subequations}}
\newcommand{\eml}{\end{subequations}}
\newcommand{\beq}{\begin{eqnarray}}
\newcommand{\eq}{\end{eqnarray}}
\newcommand{\ba}{\begin{array}}
\newcommand{\ea}{\end{array}}
\newcommand{\bpm}{\begin{pmatrix}}
\newcommand{\epm}{\end{pmatrix}}
\newcommand{\bc}{\begin{cases}}
\newcommand{\ec}{\end{cases}}
\newcommand{\lt}{\left}
\newcommand{\rt}{\right}
\newcommand{\n}{\nonumber}
\newcommand{\la}{\langle}
\newcommand{\ra}{\rangle}
\newcommand{\ep}{\varepsilon}
\newcommand{\bb}{\boldsymbol}
\newcommand{\bs}{\mathbf}
\newcommand{\h}{^\dagger}
\newcommand{\0}{^{\phantom{\dagger}}}

\DeclareMathOperator{\Hv}{\Theta}
\DeclareMathOperator{\Tr}{Tr}

\begin{document}

\title{Disorder-assisted Spin-Filtering at Metal/Ferromagnet Interfaces: An Alternative Route to Anisotropic Magnetoresistance}

\author{Ivan Iorsh}
\affiliation{Department of Physics, Engineering Physics \& Astronomy, Queen's Universiy, Kingston, Canada}

\author{Mikhail Titov}
\affiliation{Institute for Molecules and Materials, Radboud University, Nijmegen, The Netherlands}

\date{\today}

\begin{abstract}
We introduce a minimal interface-scattering mechanism that produces a sizable anisotropic magnetoresistance (AMR) in metal/ferromagnet bilayers (e.g., Pt/YIG) without invoking bulk spin or orbital Hall currents. In a $\delta$-layer model with interfacial exchange and Rashba spin-orbit coupling, charge transfer at a high-quality interface creates a spin-selective phase condition (interfacial spin filtering) that suppresses backscattering for one spin projection while enhancing momentum relaxation for the other. The resulting resistance anisotropy peaks at an optimal metal thickness of a few nanometers, quantitatively reproducing the thickness and angular dependences typically attributed to spin Hall magnetoresistance (SMR), as well as its characteristic magnitude. Remarkably, the maximal AMR scales linearly with the smaller of the two coupling strengths -- exchange or spin-orbit, highlighting a mechanism fundamentally distinct from SMR. Our scattering formulation maps onto Boltzmann boundary conditions and predicts other clear discriminants from SMR, including strong sensitivity to interfacial charge transfer and disorder.
\end{abstract}

\maketitle

The spin Hall effect (SHE) and its inverse have become cornerstones of spintronics~\cite{Hoffmann2013, Sinova2015review}. In heavy metals with strong spin-orbit coupling, an applied charge current is proposed to generate a transverse spin accumulation detectable via magnetoresistance or spin-torque phenomena. Recently, the orbital Hall effect (OHE) has been advanced as an additional channel for angular momentum transport, with theory and experiments indicating sizable orbital responses in transition metals~\cite{Go2018, Jo2018_PRB, Choi2023_Nature}. These frameworks have been widely used to interpret magnetotransport in heavy-metal/ferromagnet heterostructures, often under the umbrella of spin Hall magnetoresistance (SMR) or its orbital analogue.

Despite their success as interpretative tools, both SHE- and OHE-based pictures face conceptual ambiguities. The definition of spin or orbital current operators is not unique, and such current operators do neither correspond to conserved quantities nor couple to external fields in the respective effective Hamiltonians \cite{Rashba2003, Shi2006}. Consequently, their expectation values lack the status of genuine observables, raising questions of whether ``spin currents'' or ``orbital magnetization currents'' do necessarily provide a physically sound basis for interpreting transport experiments.

A paradigmatic case is anisotropic magnetoresistance (AMR) in metal/ferromagnet bilayers. In Pt/YIG, a system with an insulating ferromagnet, AMR is widely explained as SMR originating from reflection/absorption of spin-Hall currents at the Pt/YIG interface~\cite{Nakayama2013, Chen2013, Althammer2013, Vlietstra2013_PRB, Chen2016_review}. Related interpretations have been extended to metallic stacks (e.g., Co/Pt), where conventional AMR in the ferromagnet can coexist with SMR in the normal metal. More recently, interfacial spin-orbit magnetoresistance and Rashba--Edelstein magnetoresistance (REMR) have highlighted the role of interfacial spin-orbit scattering even without spin-current absorption~\cite{Zhou2018_SciAdv, Nakayama2016_PRL, Amin2016_Formalism, Amin2016_Phenomenology}. OHE-based scenarios, conversion of orbital transport into spin accumulation at the interface, have likewise been proposed for Pt/Co, NiFe/Pt, and Pt/YIG~\cite{DingPRL2022, SalaPRRes2022, Jo2023, Salemi2019_NatComm, Go2018, Jo2018_PRB, Choi2023_Nature}. These developments underscore the growing complexity of the field and the increasing difficulty of disentangling spin and orbital degrees of freedom.

\begin{figure}[tb]
\includegraphics[width=\linewidth]{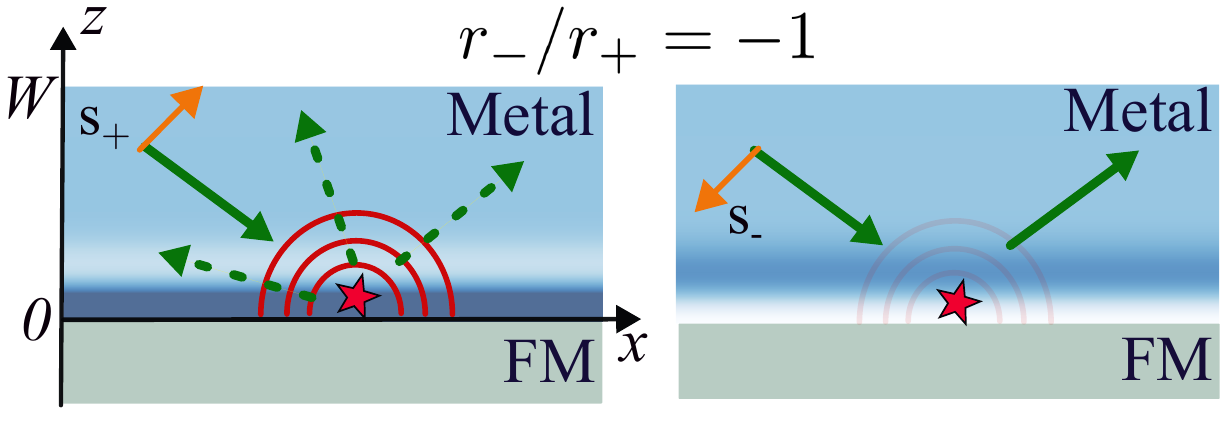}
\caption{Schematic of {\it spin filtering} for charge flow parallel to a metal/ferromagnet interface. Interfacial disorder strongly relaxes the momentum of one spin projection, whereas the other experiences nearly specular reflection (minimal momentum loss). The effect is maximized when one of the spin channels acquires interface scattering phase equal $\pi$.  \label{fig:fig1}
}
\end{figure}

In this Letter, we advance an alternative, disorder-assisted {\it spin filtering} mechanism of anisotropic magnetoresistance (SFMR) in metal/ferromagnet bilayers that does not rely on bulk spin or orbital currents. We argue that interfacial charge transfer can form a positively charged $\delta$-layer at a high-quality metal/ferromagnet interface. For suitable charge transfer, this $\delta$-layer, which is sensitive to both exchange and interfacial spin-orbit coupling (ISOC), mediates a resonant interface scattering channel akin to resonant surface/interface states in tunneling anisotropic magnetoresistance~\cite{Chantis2007_PRL, Shen2020_PRB}. The interference between ordinary non-magnetic impurity scattering and the spin-selective resonant interface channel produces strong {\it spin filtering}: electrons with one spin projection experience nearly specular reflection, while those with the opposite projection undergo enhanced momentum relaxation as shown schematicaly in Fig.~\ref{fig:fig1}. The net outcome is a robust anisotropic magnetoresistance.

A salient prediction of SFMR is that, at optimal conditions, the magnitude of the anisotropic resistance scales linearly with the interface magnetic exchange or with the ISOC strength (depending on what scale is smaller). For an ideal interface the effect is also suppressed by the metal disorder parameter $1/E_\textrm{F}\tau$, where $\tau$ is the mean scattering time in the metal bulk and $E_\textrm{F}$ is the Fermi energy. Contrary to the intuition, the effect may be enhanced by additional interfacial disorder. This naturally yields the AMR amplitude $\mathrm{\Delta}\rho/\rho$ in the $10^{-4}$--$10^{-3}$ range observed in heavy-metal films. The resulting signal is strongly enhanced near an optimal charge transfer (for Pt/YIG, approximately one electron per three Pt unit cells) and reproduces SMR/SOMR systematics, including a non-monotonic Pt thickness dependence with an optimum of a few nanometers~\cite{Vlietstra2013_PRB, Marmion2014_PRB, Althammer2013}. Crucially, in our framework the spin current is not a necessary construct~\cite{Amin2016_Formalism, Amin2016_Phenomenology, Gupta2020_PRL}. 

For definiteness, we consider a metal film of thickness $W$, occupying $0<z<W$, that is placed on a ferromagnetic dielectric for $z<0$. We neglect spin-orbit coupling in the metal bulk and adopt effective interfacial model of Amin and Stiles~\cite{Amin2016_Phenomenology},
\be
\label{model}
H=\frac{p^2}{2m} +V(\bs{r}) + U_0 \Hv(-z) + \hbar v_\textrm{F} \delta(z)\Gamma_{{\bs{p}}},
\e
where $v_\textrm{F}$ is the Fermi velocity, $V(\bs{r})$ denotes non-magnetic disorder, $\Hv(z)$ is the Heaviside theta function and $U_0$ is the spin-independent confinement, while the last term represents an interface potential,
\be
\Gamma_{\bs{p}}=u_0+\gamma\, \bb{\sigma} \cdot \hat{\bs{m}} + {\lambda}\, \bb{\sigma} \cdot ({\hat{\bs{p}}}\times\hat{\bs{z}}), \label{eq:Z}
\e
where $u_0$ sets the spin-independent barrier strength defined by the charge transfer across the interface, $\gamma$ parameterizes the interfacial exchange, and $\lambda$ quantifies the interfacial Rashba coupling; $\hat{\bs{m}}$ is the unit magnetization vector in the ferromagnet, $\hat{\bs{p}}=\bs{p}/m v_\textrm{F}$ the unit momentum vector, and $\hat{\bs{z}}$ the interface normal.

This $\delta$-layer model captures the minimal ingredients of SFMR (charge transfer, interfacial exchange and ISOC) and can be employed straightforwardly for the derivation of the scattering-matrix boundary conditions for Boltzmann transport~\cite{Amin2016_Formalism, Amin2016_Phenomenology, Shumilin2023_KineticREE, Okulov1979_SovJLowTempPhys, Wolynes1976_PRA_HydrodynamicBC, Falkovsky1983_AdvPhys_TransportSurfaces}. 

The role of interfacial disorder in SMR was recently examined in Ref.~\cite{Shumilin2023_KineticREE}. That work applies Boltzmann kinetic theory to the model of Eq.~(\ref{model}) together with the Okulov--Ustinov boundary conditions at the metal/ferromagnet interface~\cite{Okulov1979_SovJLowTempPhys}. It shows that disorder scattering at or near the interface can drive a small spin current across the interface that scales as $\lambda^3/E_\textrm{F}\tau$ \cite{Shumilin2023_KineticREE}. Within that framework the magnetoresistance vanishes; invoking the usual SHE--inverse-SHE logic~\cite{Sinova2015review} would then suggest an MR of order $\lambda^6/(E_\textrm{F}\tau)^2$ -- an exceptionally small effect.

Below we re-examine the same model and evaluate the spin current slightly away from the interface. We find that it scales as $\lambda/E_\textrm{F}\tau$ and, more importantly, that it varies by roughly six orders of magnitude within a Fermi wavelength $\lambda_\textrm{F}$ from the interface.

In contrast, allowing for finite interfacial exchange $\gamma$ in the boundary condition immediately yields a non-zero anisotropic magnetoresistance of the metal film without any appeal to spin transport. Moreover, we identify a resonant regime for $2u_0 \simeq -\sqrt{U_0/E_\textrm{F}}$, where the film magnetoresistance reaches its maximum provided $\gamma \simeq \lambda$. In this regime, for thin metal films, the resulting AMR is set by $\mathrm{\Delta}\rho/\rho\simeq \min\{|\gamma|, |\lambda|\}/E_\textrm{F}\tau$, which is naturally matching the observed magnitude of AMR~\cite{Nakayama2013, Chen2013, Althammer2013, Vlietstra2013_PRB, Chen2016_review}. For thicker films, the interfacial SFMR contribution is additionally suppressed by a factor $\ell/W$, where $\ell=v_\textrm{F}\tau$ is the mean free path.

\begin{figure}[tb]
\includegraphics[width=\linewidth]{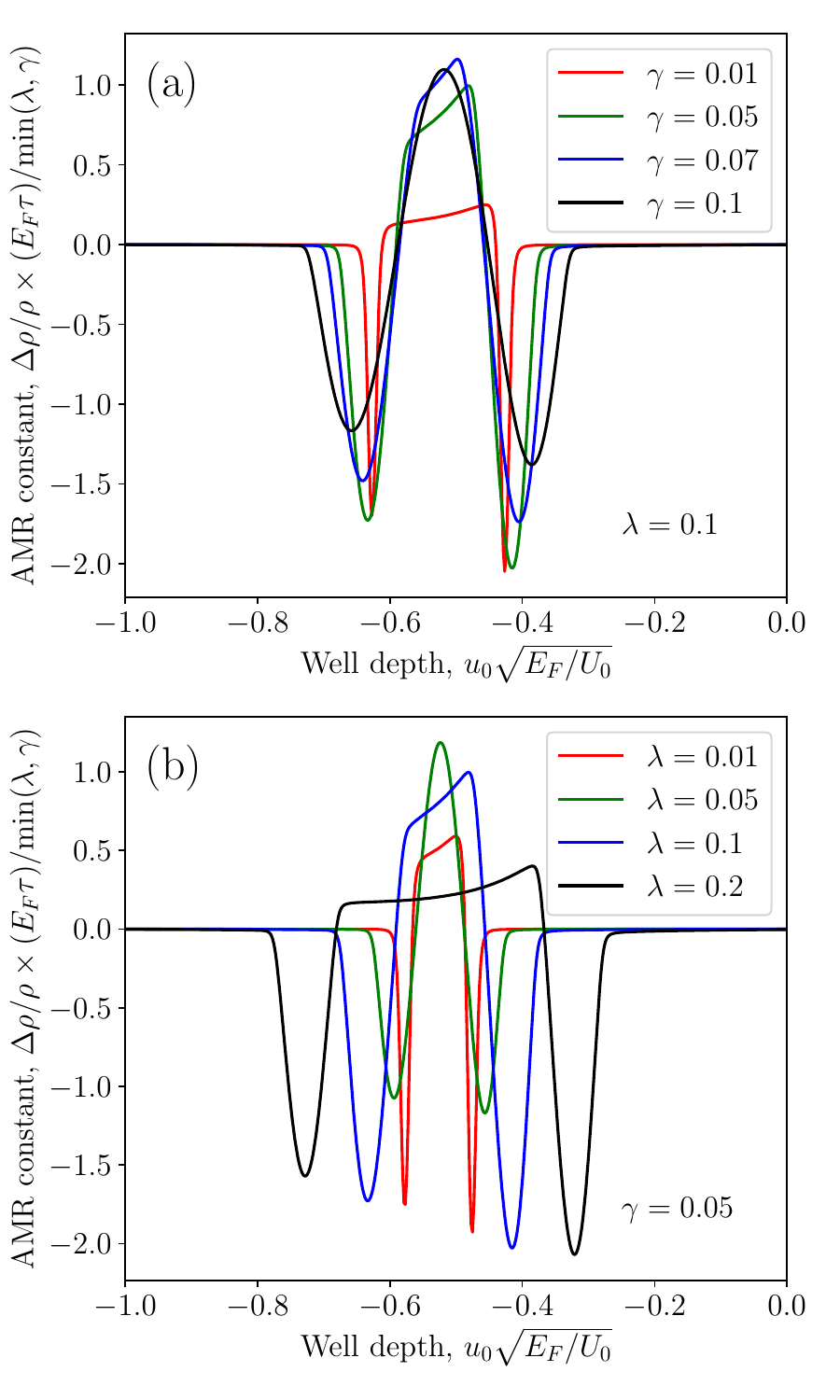}
\caption{The AMR constant $\mathrm{\Delta}\rho/\rho$ as a function of the charge transfer parameter $u_0$ for different values of the interface exchange and Rashba couplings, $\gamma$ and $\lambda$. Both figures correspond to a sufficiently narrow metal film with $W/\ell=0.5$. Large AMR signal of both signs is observed for $2u_0\simeq -\sqrt{U_0/E_\textrm{F}}$ and $\gamma\simeq \lambda$. 
\label{fig:Fig2}
}
\end{figure}

We start by writing down the semiclassical Boltzmann equation (SBE) for the distribution function $\hat{f}$ of electrons in a metal subject to an electric field $\bs{E}=E\hat{\bs{x}}$ applied along the $x$ direction. Although both electron scattering and propagation in the metal bulk are spin-independent, we must allow for a non-trivial spin structure of the electron distribution due to the spin-selective boundary condition at the metal/ferromagnet interface. 

The stationary SBE, written in the plane-wave basis with momentum $\bs{p}$, takes the form
\be
\bs{v}_{\bs{p}}\cdot\bb{\nabla}_{\bs{r}} \hat{f}(\bs{p}, \bs{r}) - e\bs{E}\cdot\bb{\nabla}_{\bs{p}} \hat{f}(\bs{p}, \bs{r}) = -\frac{\hat{f}(\bs{p}, \bs{r}) - f_0(\ep_\bs{p})}{\tau},
\label{eq:SBE}
\e
where $\bs{v}_{\bs{p}}=\bs{p}/m$ is the electron velocity, $\ep_\bs{p}=p^2/2m$ the energy dispersion, $\hat{f}$ a $2\times2$ matrix in spin space, $\tau$ the scattering time, and $f_0(\ep)$ the angle-averaged equilibrium distribution function, which is spin-independent.

We look for the solution of Eq.~(\ref{eq:SBE}) in the form 
\be
\hat{f}(\bs{p}, \bs{r}) = f_0(\ep_\bs{p}) + f_1({\bs{p}}, z) + \hat{f}_2({\bs{p}}, z),
\e 
where $f_1$ and $\hat{f}_2$ are the non-equilibrium corrections proportional to the applied electric field $E$. The non-trivial spin structure is contained only in $\hat{f}_2$, and the condition $\hat{f}_2 \ll f_1$ is assumed. 

The SBE is supplemented by two boundary conditions: one at the metal/air interface ($z=W$) and another at the metal/ferromagnet interface ($z=0$). We express $f_{1,2}$ as $f_{1,2} = f_{1,2}^{+}\Hv(p_z) + f_{1,2}^{-}\Hv(-p_z)$, distinguishing electrons moving away from and toward the interface, respectively. At $z=W$ we impose the standard Fuchs-Sondheimer boundary condition for a perfectly diffusive surface~\cite{fuchs1938conductivity, sondheimer2001mean}: $f_{1,2}^{-}({\bs{p}}, W) = 0$. At $z=0$, we assume almost specular scattering, where $\hat{f}_2$ represents deviations from perfect specularity. In this case, the boundary condition for $f_1$ is $f_{1}^{-}(\ep_\bs{p}, \hat{\bs{p}}, 0) = f_{1}^{+}(\ep_\bs{p}, \hat{\bs{p}}, 0)$, while for $\hat{f}_2$ we adopt the general Okulov-Ustinov boundary condition~\cite{okulov1979surface},
\begin{align}
&|v_z|\!\lt[ \hat{f}_2^{+}({\bs{p}},0) - \hat{f}_2^{-}({\bs{p}},0)\rt] \n\\
&=\int_{p_z'<0}\!\!\!\hat{\mathcal{W}}(\bs{p},\bs{p}')\!\lt[f_1^{-}({\bs{p}},0) - f_1^{-}({\bs{p}'},0)\rt]\frac{d^3\bs{p}'}{(2\pi)^3},
\end{align}
where $v_{\alpha}$ is the velocity component, and $\hat{\mathcal{W}}(\bs{p},\bs{p}')$ is the (yet unspecified) scattering rate. On the right-hand side we have already used that $\hat{f}_2 \ll f_1$.

The solution for $f_1$ reads:
\begin{align}
&f_1(\bs{p}, z) = eE v_\textrm{F}\tau \cos\phi_{\bs{p}}\sqrt{p^2-p_z^2}\,(\pa f_0/\pa \ep_\bs{p}) \n\\ 
&\times\lt[\left(1-e^{\frac{z-W}{|v_z|\tau}}\right)\Hv(-\hat{p}_z) + \left(1-e^{-\frac{W+z}{v_z\tau}}\right)\Hv(\hat{p}_z)\rt].
\end{align}
The solution for $\hat{f}_2$ has the form:
\be
\hat{f}_2({\bs{p}}, z) = \hat{f}_2({\bs{p}},0)\Hv(p_z)\exp(-z/v_z\tau),
\e
and its boundary value is obtained from
\be
\hat{f}_2({\bs{p}},0) = \int\displaylimits_{p_z'<0}\!\frac{d^3\bs{p'}}{(2\pi)^3} \frac{\hat{\mathcal{W}}(\bs{p},\bs{p}')}{|v_z|}\lt[f_1^-(\bs{p},0) - f_1^-(\bs{p'},0)\rt]. \n
\e
The scattering rate is given by the Fermi golden rule,
\be
\hat{\mathcal{W}}(\bs{p},\bs{p}') = 2\pi N_i V_0^2\,\hat{s}_{\bs{p}}\hat{s}_{\bs{p}'}\hat{s}^{\dagger}_{\bs{p}'}\hat{s}^{\dagger}_{\bs{p}}\,\delta(\ep_{\bs{p}}-\ep_{\bs{p}'}),
\e
where $\hat{s}_{\bs{p}} = 1 + \hat{r}_{\bs{p}}$, $N_i$ is the surface impurity concentration, and $V_0$ the impurity potential strength. Interface reflection matrix $\hat{r}_{\bs{p}}$ appear because specularly reflected electrons near the interface remain coherent with incident ones, leading to interference effects. The spin structure of $\hat{\mathcal{W}}$ reflects spin mixing during specular reflection at an interface with spin-orbit coupling.

Once $\hat{f}_2$ is known, we can compute two-dimensional charge and spin current densities using the standard thermodynamic definition 
\be
J_{\beta}^{\alpha} = \int\limits_0^{W}\!\!dz\! \int \!\!\frac{d^3\bs{p}}{(2\pi)^3}\, ev_{\beta} \Tr\lt[\sigma_{\alpha}\big(f_1(\bs{p},z)+\hat{f}_2(\bs{p},z)\big)\right].
\label{eq:Jgeneral}
\e
where $\alpha = 0, x, y, z$ is the spin index ($\alpha=0$ denotes charge), while the index $\beta=x, y, z$ refers to the velocity directions. 

The total current separates into an isotropic, spin-independent part from $f_1$ and an anisotropic, spin-dependent part from $\hat{f}_2$. The corresponding conductivity can be decomposed as $\sigma_{\beta\beta'}^{\alpha} = \sigma_{1}\delta_{\beta\beta'}\delta_{\alpha 0} + \sigma_{2,\beta\beta'}^{\alpha}$. Assuming $\sigma_1 \gg \hat{\sigma}_2$, the relative anisotropic contribution to the resistivity is $\mathrm{\Delta}\rho_{\|(\perp)}(\varphi)/\rho \approx -\sigma_{2}/\sigma_1$, where $\|$ and $\perp$ denote directions parallel and perpendicular to the electric field, respectively, and $\rho$ is the isotropic resistivity of the film. The expressions for $\mathrm{\Delta}\rho_{\|,\perp}$ yield:
\begin{align}
&\frac{\mathrm{\Delta}\rho_{\|,\perp}}{\rho} = \frac{3(E_\textrm{F}\tau)^{-1}}{8\pi^2 \mathrm{\Phi}(2w)}\frac{N_i}{N_0\lambda_F}
\int\frac{d^2{\bs{k}}\,d^2{\bs{k}'}}{qq'}k_{\|,\perp} \frac{1-e^{-\frac{w}{q}}}{w} \n\\
&\times S_{\bs{k},\bs{k}'}\!\lt[k\!\lt(1\!-\!e^{-\frac{w}{q}}\rt)\!\cos\phi_{\bs{k}}\! -\! 
k'\!\lt(1\!-\!e^{-\frac{w}{q'}}\rt)\!\cos\phi'_{\bs{k}}\rt],
\label{eq:deltarho}
\end{align}
where $N_0$ is a bulk impurity concentration, $N_i$ is a two-dimensional surface impurity concentration, $w=W/\ell$, $k_\parallel=p_x/mv_{\textrm{F}}=k\cos\phi_{\bs{k}}$, $k_\perp=p_y/mv_{\textrm{F}}=k\sin\phi_{\bs{k}}$ are the dimensionless in-plane momenta, $q=\sqrt{1-k^2}$, and
\be
S_{\bs{k},\bs{k}'} =\Tr[\hat{M}_{\bs{k}}\hat{M}_{\bs{k}'}], \qquad 
\hat{M}_{\bs{k}} = \hat{s}_{\bs{k}}\h\hat{s}\0_{\bs{k}}=\hat{s}\0_{\bs{k}}+\hat{s}_{\bs{k}}\h.
\e
The function $\mathrm{\Phi}(2w)$ describes the classical size effect in thin metallic films~\cite{parrott1965new} (see Supplementary Information for explicit form). For $w\gg 1$, $\mathrm{\Phi}(2w) \approx 1$, while for $w\ll 1$, $\mathrm{\Phi}(2w) \approx -(3w/4)\ln w$.

For a perfect interface $N_i \simeq N_0\lambda_F$ due to the bulk impurities within the distance $\lambda_\textrm{F}$ to the interface. The presence of interfacial defects ensures $N_i\gtrsim  N_0\lambda_F$.  The small parameter $1/E_\textrm{F}\tau$ indicates that bulk corrections of the same order may not be captured by SBE. This is, however, of no importance for AMR since the bulk Boltzmann equation lacks spin-dependent terms and such ``weak localization'' corrections can be safely neglected.

The angular dependence of longitudinal and transverse resistivity components is $\mathrm{\Delta}\rho_{\|}\sim \mathrm{\Delta}\rho\cos 2\varphi$ and $\mathrm{\Delta}\rho_{\perp}\sim \sin 2\varphi$. In Figs.~\ref{fig:Fig2}(a,b) we plot $\mathrm{\Delta}\rho/\rho$ versus the charge transfer parameter $u_0$ for various $\lambda$ and $\gamma$. A resonant enhancement of magnetoresistance appears within a certain range of $u_0$, originating from the structure of the matrix $\hat{M}_{\bs{k}}$:
\be
\label{MkF}
\hat{M}_{\bs{k}} = \frac{2(\mathrm{Im}\mathcal{K})^2}{||Z_{\mathbf{k}}|^2-\mathcal{K}^{2}|^2}
\bpm
|Z_{\mathbf{k}}|^2+|\mathcal{K}|^{2} & -2Z^*_{\bs{k}}\frac{\mathrm{Re}\mathcal{K}}{\mathrm{Im}\mathcal{K}} \\
-2Z_{\bs{k}}\frac{\mathrm{Re}\mathcal{K}}{\mathrm{Im}\mathcal{K}} & |Z_{\bs{k}}|^2+|\mathcal{K}|^{2}
\epm,
\e
where $Z_{\mathbf{k}} = \gamma e^{i\varphi} - i\lambda k e^{i\phi_{\mathbf{k}}}$, with $\varphi$ and $\phi_{\mathbf{k}}$ being the angles of the exchange field and momentum, respectively,  and $\mathcal{K} = \sqrt{U_0/E_\textrm{F} - q^2} + 2u_0 + iq$. 

\begin{figure}[tb]
\includegraphics[width=\linewidth]{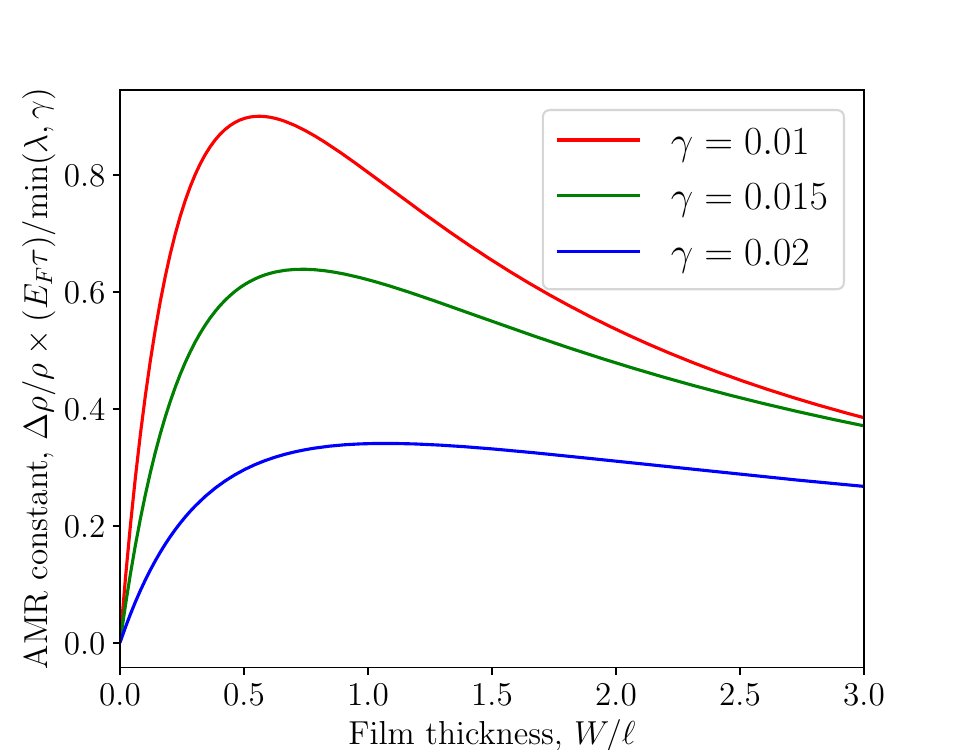}
\caption{AMR constant as a function of film thickness $W/\ell$ for three values of interface exchange parameter $\gamma$ and for $E_\textrm{F}/U_0=0.8$, $\lambda = 0.01$, $2u_0=-\sqrt{U_0/E_\textrm{F}}$. 
\label{fig:fig3}}
\end{figure}

For generic $u_0$ and $\gamma,\lambda \ll 1$, we have $|Z_{\mathbf{k}}|^2 \ll |\mathcal{K}|^2$, and the anisotropic term in Eq.~\eqref{eq:deltarho} scales as $\lambda^2\gamma^2$.

When $2u_0 \simeq -\sqrt{U_0/E_\textrm{F}}$, one may find the values of $q \in(0,1)$ that yield the condition $\mathrm{Re}\,\mathcal{K}=\pm|Z_\bs{k}|$. In this case, one of the eigenphases of the matrix $\hat{r}_\bs{k}$ equals $\pi$, while the other one remains close to zero. As the result, the denominator in Eq.~(\ref{MkF}) becomes arbitrarily small leading to a resonant enhancement of the AMR signal. In the leading order in $\lambda,\gamma$, we find:
\begin{align}
\frac{\mathrm{\Delta}\rho}{\rho} \approx \frac{\mathrm{min}(|\lambda|,|\gamma|)}{\mathrm{\Phi}(2w)\,E_\textrm{F}\tau}
\left[\frac{1}{5}+\frac{4}{5}\!\left(\frac{\mathrm{min}(|\lambda|,|\gamma|)}{\mathrm{max}(|\lambda|,|\gamma|)}\right)^{\!2}\right],
\end{align}
where $\mathrm{\Phi}(2w)$ depends only on the film thickness. At resonance, the AMR is linear in the smaller of the two parameters, $\gamma$ or $\lambda$. This condition corresponds to the situation shown in Fig.~\ref{fig:fig1}, where one spin projection becomes immune to impurity scattering -- an interface spin-filtering regime determined by $\mathrm{min}(\gamma,\lambda)$.

\begin{figure}[tb]
\includegraphics[width=\linewidth]{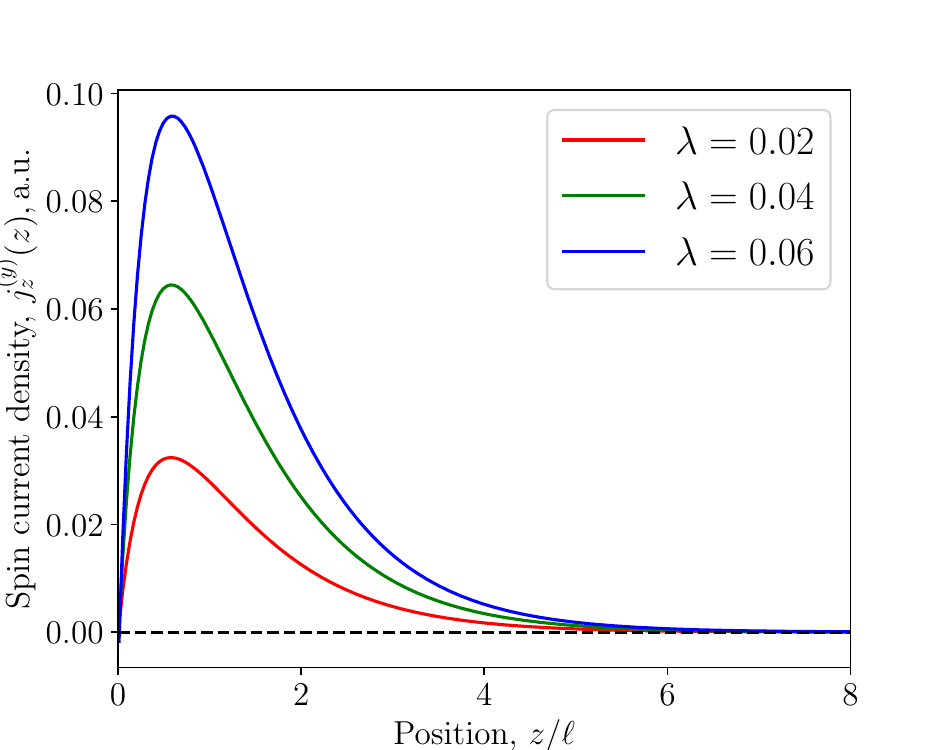}
\caption{Spatial dependence of the spin current density $\sigma_y v_z$ for vanishing exchange parameter $\gamma=0$, $u_0=-0.4$, $E_\textrm{F}/U_0=0.8$, and film thickness $W/\ell=8$. 
\label{Fig:fig4}}
\end{figure}

In Fig.~\ref{fig:fig3} we plot the dependence of the anisotropic magnetoresistance on the film thickness. It can be seen that it generally reproduces the experimentally observed phenomena with a maximum value for the thickness $W$ approximately equal to the mean free path $\ell$ with subsequent $W^{-1}$ decay for larger $W$. The suppression of the AMR in the limit $W/\ell\ll 1$ is largely due to the strong deviation of $\mathrm{\Phi}(2w)$ from one -- the classical size effect. 

Equation~\eqref{eq:Jgeneral} can also be used to define the spin current. Even for $\gamma=0$ (when AMR is forbidden by symmetry) a finite spin current with polarization along $y$ flows in the $z$ direction as shown in Fig.~\ref{Fig:fig4}. The current density demonstrates a strong dependence on the choice of $z$, indicating that the spin current is not conserved on length scales of the order of the mean free path.  At resonant conditions $2u_0\simeq -\sqrt{U_0/E_\textrm{F}}$ and away from the interface, the spin current scales as $\lambda\ln(\lambda^{-1})$ for $\lambda\ll 1$, while exactly at the interface ($z=0$) it scales as $\lambda^3\ln(\lambda^{-1})$. In contrast, the proposed spin filtering magnetoresistance scales linearly with $\lambda$ for $\lambda \ll \gamma$ and does not depend on $\lambda$ for $\lambda \gg \gamma$. Thus, it cannot be attributed to a combination of spin Hall and inverse spin Hall effects.

The microscopic origin of the exchange parameter $\gamma$ requires further study. As shown experimentally~\cite{Nakayama2013}, the AMR persists even when a thin non-magnetic spacer separates the FM and metal, suggesting that direct exchange is suppressed. An indirect exchange mediated, for example, by the RKKY interaction may therefore play the dominant role. 

In our calculations, we assumed no additional interface impurities; scattering arises from bulk impurities in a coherence layer of thickness $\lambda_\textrm{F}$ near the interface, leading to the smallness parameter $1/E_\textrm{F}\tau$. Increasing $N_i$ enhances the effect, and in the limit $N_i \simeq \ell^{-2}$ the factor $1/E_\textrm{F}\tau$ cancels out, leaving $\mathrm{min}(\lambda,\gamma)$ as the only small parameter. This indicates that careful interface engineering and control over the charge transfer parameter $u_0$ can, in principle, enhance the AMR by orders of magnitude.

In conclusion, we have proposed a new mechanism for anisotropic magnetoresistance in FM/NM bilayers that does not rely on the spin Hall or inverse spin Hall effect phenomenology but rather on anisotropic interfacial electron scattering. The effect requires no bulk spin-orbit coupling, yet reproduces the correct magnitude, angular dependence, and film-thickness behavior observed experimentally. These results underscore the importance of quantitatively accurate boundary conditions for a proper description of magnetoresistance in FM/NM bi-layers.

We appreciate discussions with Joseph Barker, Sebastian Goennenwein, Igor Gornyi and Yuriy Mokrousov.

\bibliographystyle{apsrev4-2}
\bibliography{sample}

\supplementarystart

\begin{center}
\bfseries\large SUPPLEMENTARY INFORMATION\\[6pt]
\bfseries\large Disorder-assisted Spin-Filtering at Metal/Ferromagnet Interfaces: An Alternative Route to Anisotropic Magnetoresistance\\[6pt]
Ivan Iorsh and Mikhail Titov
\end{center}

\begin{quote}
In this supplementary we provide the details of the derivations for an interested reader.
\end{quote}

\section{Derivation of the elastic reflection matrix}

We start from the Hamiltonian of Eq.~(1) from the main text,
\be
H=\frac{p^2}{2m} +V(\bs{r}) + U_0\Hv(-z) + \hbar v_\textrm{F} \delta(z) \Gamma_{\bs{p}},
\e
where $v_\textrm{F}$ is the Fermi velocity, $V(\bs{r})$ denotes nonmagnetic (Coulomb) disorder in the metal bulk, and $U_0\Hv(-z)$ is the confinement potential for conduction electrons. Here $U_0$ has a meaning of the work function (the energy needed for an electron to leave the metal for a ferromagnet dielectric at $z<0$). 

The $\delta$-layer potential is set by 
\be
\Gamma_{\bs{p}}=u_0+\gamma\, \bb{\sigma} \cdot \hat{\bs{m}} + \lambda\, \bb{\sigma} \cdot (\hat{\bs{p}}\times\hat{\bs{z}})
=\bpm u_0 +\gamma\cos\theta & Z_{\bs{p}}^* \\ Z_{\bs{p}} & u_0 - \gamma\cos\theta \epm.
\e
where $\hat{\bs{p}}=\bs{p}/mv_\textrm{F}=(k_x,k_y, q)$,  $\hat{\bs{m}}=(\sin\theta \cos\varphi,\sin\theta\sin\varphi, \cos\theta)$ is the unit vector in magnetization direction of the ferromagnet, and $\hat{\bs{z}}$ is the unit vector normal to the interface. Note that the operator $\Gamma_{\bs{p}}$ is commuting with $\delta(z)$. 

Here, the dimensionless parameters $u_0$, $\gamma$ and $\lambda$ control the interface charge transfer, effective exchange interaction and the Rashba interface spin-orbit coupling, respectfully. The absence of either spin-orbit coupling or exchange would immediately lead to an identically vanishing AMR on symmetry grounds. 
  
In a plain-wave eigenstate the operator $\Gamma_{\bs{p}}$ is replaced by the momentum-dependent matrix function, where  
\be
{Z}_{\bs{p}}=\gamma \sin\theta e^{i\varphi}-i\lambda k e^{i\phi_{\bs{k}}}, \qquad \mbox{where}\quad k=\sqrt{1-q^2},
\e
and $\phi_{\mathbf{k}}$ is the angle of in-plane electron momentum, $k_x=k\cos\phi_{\bs{k}}$, $k_y=k\sin\phi_{\bs{k}}$.

The reflection matrix is defined from the scattering problem for the interface reflection in a clean system $V(\mathbf{r})=0$. For elastic scattering we fix the conduction electron energy to $E_\textrm{F}=mv_\textrm{F}^2/2$ and use the dimensionless coordinate $\zeta= m v_\textrm{F} z$. 

In this case the scattering problem ($\hbar=1$) for the propagation in $z$ direction is reduced to the spectral equaiton
\be
\label{spec}
\lt[-\pa^2_\zeta +2 \Gamma_{\bs{k}} \delta(\zeta)\rt] \Psi_{\bs{k}} =\lt[1-k^2 - U_0 \Hv(-\zeta)/E_\textrm{F}\rt]\Psi_{\bs{k}},
\e
where we can also use $q=\sqrt{1-k^2}$. 

The equation of Eq.~(\ref{spec}) is solved by 
\be
\Psi_{\bs{k}}(\zeta>0)= (a_{\bs{k}} e^{-i q \zeta} + b_{\bs{k}} e^{i q \zeta})/\sqrt{q},\qquad 
\Psi_{\bs{k}}(\zeta<0) = c_{\bs{k}} e^{\kappa \zeta}/\sqrt{q},
\e
where $q$ is defined as positive definite, $\kappa=\sqrt{U_0/E_\textrm{F}-q^2}$, $a_{\bs{k}}$, $b_{\bs{k}}$ and $c_{\bs{k}}$ are spinors and the factor $1/\sqrt{q}$ ensures the correct normalization of the scattering state. 

The reflection matrix is obtained from the relation $b_{\bs{k}} = \hat{r}_{\bs{k}} a_{\bs{k}}$ using matching conditions at $\zeta=0$, which read 
\be
\lt.\Psi_{\bs{k}}(\zeta<0)\rt|_{\zeta=0} = \lt.\Psi_{\bs{k}}(\zeta>0)\rt|_{\zeta=0}, \qquad \lt[\frac{\pa\Psi_{\bs{k}}(\zeta>0)}{\pa \zeta}-\frac{\pa\Psi_{\bs{k}}(\zeta<0)}{\pa \zeta}\rt]_{\zeta=0} = 2\Gamma_{\bs{k}}\Psi_{\bs{k}}(\zeta=0).
\e
Thus, we get $c_{\bs{k}}=(1+\hat{r}_{\bs{k}}) a_{\bs{k}}$ and $(2\Gamma_{\bs{k}}+\kappa)c_{\bs{k}}=-iq(1-\hat{r}_{\bs{k}}) a_{\bs{k}}$, which gives
\be
\frac{1-\hat{r}_{\bs{k}}}{1+\hat{r}_{\bs{k}}} = i \frac{2\Gamma_{\bs{k}}+\kappa}{q}.
\e
The reflection matrix is, therefore, given by
\be
\hat{r}_{\bs{k}} = \frac{1- i K}{1+i K}= - \frac{2\Gamma_{\bs{k}}+\kappa+ i q}{2\Gamma_{\bs{k}}+\kappa-i q}, \qquad \mbox{where}\quad K=\frac{2\Gamma_{\bs{k}}+\kappa}{q}.
\e
The eigenvalues of the matrix $K$ can be written as
\be
K_\pm=\frac{1}{q}\lt(2u_0+\kappa\pm 2 \sqrt{\gamma^2+\lambda^2k^2+2\lambda k \gamma\sin\theta\sin(\phi_{\bs{k}}-\varphi)}\rt).
\e
The corresponding eigenvalues of the reflection matrix $\hat{r}_{\bs{k}}$ are often parameterized with the scattering phases, $\delta_\pm$, as $r_\pm = \exp(i\delta_\pm)$.
The latter are clearly defined by 
\be
\delta_\pm = -i 
\ln\lt(
\frac{u_0+\kappa/2 \pm \sqrt{\gamma^2+\lambda^2k^2+2\lambda k \gamma\sin\theta\sin(\phi_{\bs{k}}-\varphi)}+iq/2}
{u_0+\kappa/2 \pm \sqrt{\gamma^2+\lambda^2k^2+2\lambda k \gamma\sin\theta\sin(\phi_{\bs{k}}-\varphi)}-iq/2}
\rt)
\e
For realistic interfaces we normally have $\gamma \ll \lambda \ll 1$. This means that, for an interface with $u_0+\kappa/2 \gg 1$, both scattering phases $\delta_\pm$ are identical and close to zero. However, for an interface with an optimal charge transfer $u_0+\kappa/2 \ll \gamma$ (i.\,e. for negative $u_0$ that is close to $- \sqrt{U_0/E_\textrm{F}}/2$) one may expect a large difference between the scattering phases for two spin components. In this case one may find incident waves (the values of $q$) for which one of the scattering phases equals $\pi$, i.e. corresponds to a strong resonant scattering at the interface. In the presence of disorder near the interface this will lead to a diffusive boundary condition for one spin component, while the other spin component will be specularly reflected, thus strongly affecting the AMR of the entire metal film. We refer to this effect as the interfacial spin filtering.

The interfacial spin filtering takes place when either $\delta_+$ or $\delta_-$ turns into $\pi$ for a certain value of $q$ within the range $0<q<1$. As the result, the AMR as the function of $u_0$ demonstrates two corresponding distinct peaks illustrated in Fig.~2 of the main text.  

\section{Calculation of the AMR and spin current}

The stationary SBE takes the form
\be
\bs{v}_{\bs{p}}\cdot\bb{\nabla}_{\bs{r}} \hat{f}(\bs{p}, \bs{r}) - e\bs{E}\cdot\bb{\nabla}_{\bs{p}} \hat{f}(\bs{p}, \bs{r}) =-\frac{\hat{f}(\bs{p}, \bs{r}) - f_0(\ep_\bs{p})}{\tau}, 
\label{eq:SBE_sup}
\e
where $\bs{v}_{\bs{p}}=\bs{p}/m$ is the electron velocity, $\ep_\bs{p}=p^2/2m$ is the dispersion relation, $\hat{f}$ is a matrix in spin space, $\tau$ is the scattering time, and $f_0(\ep)$ is the momentum angle-averaged distribution function, which is a unit matrix in spin space.  The SBE is subject to two boundary conditions: at $z=0$ metal/ferromagnet interface and $z=W$ metal/air interface. We assume completely diffuse scattering at $z=W$ interface, and almost specular reflection at $z=0$. Strong deviations from the specular reflection for one of the spin components due to the resonant spin-dependent scattering off metal/ferromagnet interface is responsible for a strong AMR in or model. At optimal conditions the only smallness of the AMR signal comes from an effective exchange parameter $\gamma$ (assuming $\gamma\ll \lambda$). Finite symmetry braking parameter $\gamma$ is, in any case, required in any model of the effect. 

We look for the solution of Eq.~(\ref{eq:SBE_sup}) in the form 
\be
\hat{f}(\bs{p}, \bs{r})=f_0(\ep_\bs{p})+f_1(\ep_\bs{p}, \hat{\bs{p}},z)+\hat{f}_2(\ep_\bs{p},\hat{\bs{p}}, z),
\e 
where $f_1(\ep_\bs{p}, \hat{\bs{p}},z)$ is the spin-independent angular harmonic of the distribution function that defines the main (spin-independent) contribution to the charge current, while 
$\hat{f}_2(\ep_\bs{p},\hat{\bs{p}}, z)$ is the smaller spin-dependent correction ($f_2\ll f_1$) that is a matrix in spin space and is sensitive to magnetization direction $\hat{\bs{m}}$ in the ferromagnet. The correction $\hat{f}_2$ originates from the scattering off the high-quality metal/ferromagnet interface. 

For $f_1$ we employ the standard Fouks boundary conditions. Those relate the distribution functions of electrons coming to the interface and electrons leaving the interface. In what follows we formally decompose: $f_1=f_1^++f_1^-$ and introduce the distributions functions $f_1^{\pm}(\ep_{\bs{p}},\hat{\bs{p}},z)$, where ``$+$'' corresponds to $\hat{p}_z>0$ and ``$-$'' corresponds to $\hat{p}_z<0$.  

The spin-independent harmonics $f_1$ is, then, set by the boundary conditions
\beml
\label{boundary}
\begin{align}
\label{diffusive}
&f_1^{-}(\ep_\bs{p}, \hat{\bs{p}},W)=0,\\
&f_1^{+}(\ep_\bs{p}, \hat{\bs{p}},0)=f_1^{-}(\ep_\bs{p}, \hat{\bs{p}},0).
\end{align}
\eml
These boundary conditions correspond to purely diffusive scattering at $z=W$ interface, and specular reflection at $z=0$ interface.

The solution for $f_1$ to the linear order in the electric field $E$ takes the form
\be
\label{fff1}
f_1(\ep_\bs{p}, \hat{\bs{p}},z)= eE v_\textrm{F}\tau \cos\phi_{\bs{p}} \frac{\pa f_0(\ep_\bs{p})}{\pa \ep_\bs{p}} 
\lt[\lt(1-e^{\frac{z-W}{|v_z|\tau}}\rt)\Hv(-\hat{p}_z)+\lt(1-e^{-\frac{z+W}{v_z\tau}}\rt)\Hv(\hat{p}_z)\rt],
\e
where $v_z=p_z/m$ and $\phi_\bs{p}$ is the angle of the in-plane projection of the vector $\bs{p}$ with respect to $x$ direction (this is equivalent to $\phi_{\bs{k}}$ of the previous Section).  

The spin-dependent part of the distribution function $\hat{f}_2$ yields $f_2^{-}(\ep_\bs{p}, \hat{\bs{p}},z)=0$ due to the diffusive boundary condition at $z=W$ as in Eq.~(\ref{diffusive}). At $z=W$ it yields, however, the most general form of the boundary condition from Ref.~\cite{okulov1979surface}, that can be written as
\be
\label{eq:bulk}
\hat{f}_2(\ep_\bs{p}, \hat{\bs{p}},z=0)=\frac{\Hv(p_z)}{v_z} \int \frac{d^3\bs{p}'}{(2\pi)^3} \hat{\mathcal{W}}(\bs{p},\bs{p}')\lt(f_1^{-}(\ep_\bs{p}, \hat{\bs{p}}',0)-f_1^{-}(\ep_\bs{p}, \hat{\bs{p}},0)\rt), 
\e
where the interface collision rate is set by
\be
\hat{\mathcal{W}}(\bs{p},\bs{p}') ={2\pi N_i V_0^2}\, (1+\hat{r}\0_{\bs{p}})(1+\hat{r}\0_{\bs{p}'})(1+\hat{r}\h_{\bs{p}'})(1+\hat{r}\h_{\bs{p}})\delta(\ep_{\bs{p}}-\ep_{\bs{p}'}).
\e
The coordinate dependence of the spin-dependent component is simply found as
\be
\label{f2z}
\hat{f}_2(\ep_\bs{p}, \hat{\bs{p}},z)=\hat{f}_2(\ep_\bs{p}, \hat{\bs{p}},z=0) e^{-z/v_z\tau}.
\e
Thus, the analysis of the AMR is mostly reduced to the integration in Eq.~(\ref{eq:bulk}). 

For a sake of generality we compute the charge current density alongside with the spin-current densities using the standard thermodynamic definition 
\be
\label{eq:Jgeneral_sup}
j_{\beta}^{(\alpha)}(z) =  \int \frac{d^3\bs{p}}{(2\pi)^3}\, ev_{\beta} \Tr\lt[\sigma_{\alpha}\lt(f_1(\ep_\bs{p},\hat{\bs{p}},z) +\hat{f}_2(\ep_\bs{p},\hat{\bs{p}},z)\rt)\rt]
= j\0_1(z) \delta_{\alpha,0} \delta_{\beta x} + j_{2,\beta}^{(\alpha)}(z),
\e
where $\alpha = 0,x,y,z$ is the spin index ($0$ corresponds to the usual charge current), while the index $\beta=x, y, z$ denotes the spatial directions (velocity directions). 

Since the metal layer is usually rather thin it is also convenient to define the two-dimensional current density as 
\be
J_{\beta}^{(\alpha)}=\int_0^{W} \!\!dz\; j_{\beta}^{(\alpha)}(z)=J\0_1 \delta_{\alpha,0} \delta_{\beta x} + J_{2,\beta}^{(\alpha)}.
\e
The current $J_1$ is associated with the spin-independent angular harmonics of the distribution function $f_1$. Due to the absence of spin structure in this harmonic it is simply the cvharge current flowing in the direction of the field $E$ ($x$-direction). 

With the help of the solution for $f_1$ from Eq.~(\ref{fff1}) we obtain
\begin{align}
&J_1=\sigma_0^{\textrm{2D}} \mathrm{\Phi} (2 W/l)\,E=\sigma_1 E,\\
&\mathrm{\Phi}(x) = 1-\frac{3}{8x}+ \frac{e^{-x}}{16x}\left(x^3-x^2-10x+6\right) +\frac{\mathrm{Ei}(1,x)}{16 x}(-x^4+12x^2),
\end{align}
where $\mathrm{Ei}$ is the exponential integral and $\sigma_0^{\textrm{2D}}= n_\textrm{F}e^2 \tau /m W$ is the two-dimensional Drude conductivity of the film without an effect of the boundaries (here $n_\textrm{F}$ is the three dimensional electron concentration). 

The function $\mathrm{\Phi}(x)$ was first introduced by Sondheimer~\cite{sondheimer2001mean} to describe the classical size effects in the conduction of thin metal films. 
For $x \ll 1$ one finds $\mathrm{\Phi}(x)\approx (3x/4) \ln(x^{-1})$, while for $x\gg 1$, one obtains $\mathrm{\Phi}(x)\approx 1- 3/8x$. 

The contribution $J_{2,x}^{(0)}$ is sensitive to the direction of $\hat{\bs{m}}$ and defines anisotropic resistance of the metal film. From Eqs.~(\ref{eq:bulk}, \ref{f2z}) we obtain
\begin{align}
J_{2,x}^{(0)}&=\sigma_0^\textrm{2D} E   \frac{3}{16 \pi^2} \frac{1}{E_\textrm{F}\tau} \frac{N_i}{N_0\lambda_\textrm{F}} \frac{\ell}{W} \;\int_0^1\!\!\! dk\, \int_0^{2\pi}\!\!\! d\phi_\bs{k}\, \frac{k}{q}\,k\cos\phi_{\bs{k}} \lt(1-e^{-W/\ell q} \rt) \n\\
&\times \int_0^1\!\!\! dk'\,  \int_0^{2\pi}\!\!\! d\phi'_\bs{k}\, \frac{k'}{q'}\, S_{\bs{k},\bs{k}'}\,\lt[k'\cos\phi'_{\bs{k}}\lt(1-e^{-W/\ell q'}\rt)-k\cos\phi_{\bs{k}}\lt(1-e^{-W/\ell q}\rt)\rt],
\label{eq:jy}
\end{align}
where $\lambda_\textrm{F}$ is the Fermi wave length, $N_i$ is the two-dimensional concentration of the interfacial defects, $N_0$ is the three dimensional concentration of non-magnetic impurities in the metal bulk,  $S_{\bs{k},\bs{k}'}= \Tr[(\hat{s}\0_{\bs{k}}+\hat{s}\h_{\bs{k}})(\hat{s}\0_{\bs{k}'}+\hat{s}\h_{\bs{k}'})]$, $q=\sqrt{1-k^2}$, and $q'=\sqrt{1-(k')^2}$. Here we have defined $\hat{s}_{\bs{k}}=1+\hat{r}_\bs{k}$. 
 
For a perfect interface one finds $N_i=N_0\lambda_F$. The interface disorder, in this case, is defined by ``bulk'' non-magnetic impurities located within a distance $\lambda_\textrm{F}$ from the interface. Normally, interface brings up an additional scattering hence even for a high-quality interfaces one finds $N_i/ N_0\lambda_F \gg 1$. Thus, for $W\simeq \ell$ the pre-factor
$(1/E_\textrm{F}\tau) (N_i/N_0\lambda_\textrm{F}) (\ell/W)$ in Eq.~(\ref{eq:jy}) may still be of the order of $1$.   

The charge current $J_{2,x}^{(0)}$ contains the angle-dependent contribution that varies as $\cos 2\varphi$ with respect to the in-plane direction of $\hat{\bs{m}}$. This angle-dependent part defines the AMR and can be quantified by the ratio
\begin{align}
\mathrm{\Delta}\rho_\parallel/ \rho = -\lt[J_{2,x}^{(0)}-\lt\la J_{2,x}^{(0)}\rt\ra_{\varphi}\rt]/{J_1},
\end{align}
where $\lt\la J_{2,x}^{(0)}\rt\ra_{\varphi}\ll J_1$ is the part of $J_{2,x}^{(0)}$ that is independent of $\varphi$ (the angular brackets $\la\dots\ra_\varphi$ denote the averaging over the directions of $\varphi$). 

Similarly one can define the tranversal current contribution $J_{2,y}^{(0)}$, which is proportional to $\sin 2\varphi$. The coefficient in $J_{2,x}^{(0)}$, which is proportional to $\cos2\varphi$ and the coefficient in $J_{2,y}^{(0)}$ in the front of $\sin 2\varphi$ are identical in our theory and represent the interfacial AMR that is consistent with that observed in SMR experiments. We will demonstrate this point explicitly in the next Section.  

It is also instructive to see that the physics of spin-filtering magnetoresistance (SFMR) that we have presented has nothing to do with the mechanism of the SHE. This can be seen most straightforwardly within the same model by computing the corresponding spin-current density. The spin Hall effect is usually associated in this geometry with the spin-current density component $j_z^{(y)}/e$ in the metal bulk. The latter is found from the same SBE as 
\begin{align}
\frac{j_z^{(y)}(z)}{e}=&\frac{\sigma_0 E}{e} \frac{3}{16\pi^2} \frac{1}{E_\textrm{F}\tau} \frac{N_i}{N_0\lambda_\textrm{F}}  
\int_0^1 \frac{k dk}{q} e^{-z/\ell q} \int_0^{2\pi} \!\!\!d\phi_\bs{k} \, \int_0^1 \frac{k'dk'}{q'} \int_0^{2\pi} \!\!\!d\phi'_\bs{k}\, S^{(y)}_{\bs{k},\bs{k'}}\n\\
&\times \lt[k'\cos\phi'_{\bs{k}}\lt(1-e^{-W/\ell q'}\rt)-k\cos\phi_{\bs{k}}\lt(1-e^{-W/\ell q}\rt)\rt],
\end{align}
where $\sigma_0=n_\textrm{F} e^2 \tau/m$ and $S^{(y)}_{\bs{k},\bs{k'}}= \mathrm{Tr}[\sigma_y\hat{s}_{\mathbf{k}}(\hat{s}_{\mathbf{k'}}+\hat{s}^{\dagger}_{\mathbf{k'}})\hat{s}_{\mathbf{k}}^{\dagger}]$. The quantity $S^{(y)}_{\bs{k},\bs{k'}}$ brings an additional smallness in the spin-orbit strength that makes the transversal spin current, and consequently Hall angle, several magnitudes smaller than the relative AMR. As the result, the SMR, which is formally given by the squared Hall angle, is absolutely negligible as compared to the SFMR even outside the optimal parameter range $u_0\simeq - \sqrt{U_0/E_\textrm{F}}$. Interestingly, the value of the spin current at the interface $j_z^{(y)}(z=0)$ scales as $\lambda^3$ and is practically vanishing for any realistic values of the parameters. 

\section{Angular dependence of the interfacial AMR signal}

In order to illuminate the dependence of the AMR on the in-plane magnetization angle $\varphi$ analytically, we have to make few simplifications in Eq.~(\ref{eq:jy}). 
The entire dependence on $\varphi$ arises exclusively from the interface scattering kernel
\be
S_{\bs{k},\bs{k}'}= \Tr\lt[(\hat{s}\0_{\bs{k}}+\hat{s}\h_{\bs{k}})(\hat{s}\0_{\bs{k}'}+\hat{s}\h_{\bs{k}'})\rt]=
\Tr \lt[\hat{M}_{\bs{k}}\hat{M}_{\bs{k}'}\rt],
\e
where $\hat{M}_{\bs{k}}=\hat{s}\0_{\bs{k}}+\hat{s}\h_{\bs{k}}$. In a vicinity of the spin filtering resonance that dominates the AMR signal, the leading contribution comes from the product of the diagonal elements of the matrix $\hat{M}_\bs{k}$. Furthermore, in order to simplify the logic, we may neglect the thickness dependence (assuming $W>l$ and omitting the exponentials $\exp\lt(-W/\ell q\rt)$). We also may notice that the result is dominated by $q\ll 1$, hence, for an estimate we may let $k=1$ in the integrand. After that, the integration over $k$ and $k'$ is performed trivially with the result 
\begin{align}
\label{xx}
j_{2,x}^{(0)}&= A\lt[\int_0^{2\pi}\!\!\! d\phi_\bs{k} \cos^2\!\phi_\bs{k} \lt(\frac{2}{3}-\frac{\pi|Z_{\bs{k}}|}{2} \rt)\rt]\,\lt[\int_0^{2\pi}\!\!\! d\phi'_\bs{k} \lt(1-\frac{\pi |Z_{\bs{k}'}|}{2}\rt)\rt] \n\\ 
&+B\lt[\int_0^{2\pi}\!\!\! d\phi_\bs{k} \cos\phi_\bs{k} \lt(1-\frac{|Z_{\bs{k}}|}{2}\rt)\rt]\, \lt[\int_0^{2\pi}\!\!\! d\phi'_\bs{k}\cos\phi'_\bs{k}\,\lt(1-\frac{|Z_{\bs{k}'}|}{2}\rt)\rt],
\end{align}
where $A$ and $B$ are some numerical coefficients that we do not need to specify.  

Similarly, the expression for the $y$ component of the charge current can be simplified in the same approximation as the following
\begin{align}
\label{xy}
j_{2,y}^{(0)}&= A\lt[\int_0^{2\pi}\!\!\! d\phi_\bs{k} \cos\phi_\bs{k} \sin\phi_\bs{k} \lt(\frac{2}{3}-\frac{\pi|Z_{\bs{k}}|}{2} \rt)\rt]\,\lt[\int_0^{2\pi}\!\!\! d\phi'_\bs{k} \lt(1-\frac{\pi |Z_{\bs{k}'}|}{2}\rt)\rt] \n\\ 
&+B\lt[\int_0^{2\pi}\!\!\! d\phi_\bs{k} \sin\phi_\bs{k} \lt(1-\frac{|Z_{\bs{k}}|}{2}\rt)\rt]\, \lt[\int_0^{2\pi}\!\!\! d\phi'_\bs{k}\cos\phi'_\bs{k}\,\lt(1-\frac{|Z_{\bs{k}'}|}{2}\rt)\rt],
\end{align}
In the expressions of Eqs.~(\ref{xx}, \ref{xy}) we have defined 
\be
|Z_\bs{k}|=\sqrt{\gamma^2+\lambda^2+2\gamma\lambda \sin(\phi_\bs{k}-\varphi)}.
\e
Since we have to consider both $\gamma\ll 1$ and $\lambda\ll 1$, we have $|Z_\bs{k}|\ll 1$. It is easy to see that the expressions in front of the coefficient $A$ in both Eq.~(\ref{xx}) and Eq.~(\ref{xy}) contain linear terms in $|Z_\bs{k}|$, while the expressions in front of the coefficient $B$ contain only quadratic in $|Z_\bs{k}|$ terms, which we can, therefore, neglect. 

Subtracting the isotropic part from $j_{2,x}^{(0)}$, we simply obtain
\beml
\begin{align}
j_{2,x}^{(0)} - \lt\la j_{2,x}^{(0)}\rt\ra_\varphi &=- \frac{\pi^2A}{2}\int d\phi_\bs{k}\, \cos2\phi_\bs{k}\, |Z_{\mathbf{k}}|, \\
j_{2,y}^{(0)} &=- \frac{\pi^2A}{2}\int d\phi_\bs{k}\, \sin2\phi_\bs{k}\, |Z_{\mathbf{k}}|,
\end{align}
\eml
where we have neglected all terms of the order of $|Z_{\mathbf{k}}|^2$. 

We can now define $\alpha=\phi_\bs{k}-\varphi$ and take advantage of the identity
\begin{align}
\int_0^{2\pi}\!\!\! d\alpha\, \sin 2\alpha\,\sqrt{\gamma^2+\lambda^2+2\gamma\lambda \sin\alpha} = 0. 
\end{align}
As the result we obtain two anisotropic components of the charge current as
\be
\label{ANG}
j_{2,x}^{(0)}- \lt\la j_{2,x}^{(0)}\rt\ra_\varphi = \delta j\, \cos 2\varphi, \qquad j_{2,y}^{(0)} = \delta j\, \sin 2\varphi,
\e
where $\delta j$ is a common constant. 

To ensure appreciable deviations from the angular dependence of Eq.~(\ref{ANG}) one needs to make both interfacial parameters $\gamma$ and $\lambda$ of the order of $1$, which is hardly possible for any known interface.

\end{document}